\documentclass[letterpaper,12pt]{article}
\usepackage{array}
\usepackage{dcolumn}
\usepackage{float}
\usepackage[numbers]{natbib}
\usepackage{graphicx}
\usepackage{color}
\usepackage{amsmath}
\usepackage{amsfonts}
\usepackage{amsopn}
\usepackage{amsthm}
\usepackage{lscape}
\usepackage{multirow}
\theoremstyle{definition}

\theoremstyle{remark}

\DeclareMathOperator{\pr}{P}

\DeclareMathOperator{\logit}{logit}
\DeclareMathOperator{\expit}{expit}

\newcommand{\bmp}{\boldsymbol p}%
\newcommand{\bmtheta}{\mbox{\boldmath${\theta}$}}%

\renewcommand{\baselinestretch}{1.8}

\begin{document}
\begin{center}
{\LARGE A Likelihood Perspective on Dose-Finding Study Designs in Oncology}
\end{center}

\vspace{0.3cm}

\begin{center}
Zhiwei Zhang\\
Biostatistics Innovation Group, Gilead Sciences, Foster City, California, USA\\
Zhiwei.Zhang6@Gilead.com
\end{center}

\vspace{0.3cm}

\centerline{\bf Abstract}
Dose-finding studies in oncology often include an up-and-down dose transition rule that assigns a dose to each cohort of patients based on accumulating data on dose-limiting toxicity (DLT) events. In making a dose transition decision, a key scientific question is whether the true DLT rate of the current dose exceeds the target DLT rate, and the statistical question is how to evaluate the statistical evidence in the available DLT data with respect to that scientific question. This article introduces generalized likelihood ratios (GLRs) that can be used to measure statistical evidence and support dose transition decisions. Applying this approach to a single-dose likelihood leads to a GLR-based interval design with three parameters: the target DLT rate and two GLR cut-points representing the levels of evidence required for dose escalation and de-escalation. This design gives a likelihood interpretation to each existing interval design and provides a unified framework for comparing different interval designs in terms of how much evidence is required for escalation and de-escalation. A GLR-based comparison of commonly used interval designs reveals important differences and motivates alternative designs that reduce over-treatment while maintaining MTD estimation accuracy. The GLR-based approach can also be applied to a joint likelihood based on a nonparametric (e.g., isotonic regression) model or a parametric model. Simulation results indicate that the isotonic GLR performs similarly to the single-dose GLR but the GLR based on a parsimonious model can improve MTD estimation when the underlying model is correct.

\vspace{.5cm}
\noindent{Key words:}
dose de-escalation; dose escalation; dose selection; dose transition; generalized law of likelihood; isotonic regression; monotonicity; phase 1 trial design

\section{Introduction}\label{intro}

Clinical research on novel therapeutic agents in oncology usually starts with dose-finding studies, which aim to identify one or more promising doses for further evaluation. A major focus in such studies is the frequency of dose-limiting toxicity (DLT) events and the maximum tolerated dose (MTD), defined as the highest dose among those considered that has a DLT rate not exceeding the maximum tolerable DLT rate (also known as the target DLT rate). The MTD is of particular interest when drug activity is expected to increase with increasing dose and when early measures of activity are unavailable or unreliable.

Popular designs for finding the MTD include the well-known 3+3 design \citep{s89,k94}, the continual reassessment method (CRM) \citep{o90}, the accelerated titration design \citep{s97}, escalation with overdose control \citep{b98}, the Bayesian logistic regression method \textcolor{blue}{(BLRM)} \citep{n08}, the modified toxicity probability interval (mTPI) design \citep{j10}, the toxicity equivalence range (TEQR) design \citep{bl11}, the Bayesian optimal interval (BOIN) design \citep{ly15}, the i3+3 design \citep{lwj20}, and many others. Many of these designs include an up-and-down dose transition rule that assigns a dose to each cohort of patients based on accumulating DLT data, and an MTD estimation procedure to be applied to all available data at the end of the study. For interval designs \citep{j10,bl11,ly15,lwj20}, the dose transition rule amounts to locating the observed DLT rate at the current dose in one of three intervals corresponding to three possible actions (escalate, de-escalate, or stay).

In making a dose transition decision, a key scientific question to consider is whether the true DLT rate of the current dose exceeds the target DLT rate. If the right answer to this question is known with certainty, the right action to take would be obvious (de-escalate if the answer is yes; escalate if the answer is no). In reality, the right answer is unavailable, and one has to rely on statistical evidence in the available data to make a decision. One may decide to escalate or de-escalate if the evidence is interpreted as providing adequate support for such an action. If there is not enough evidence to support either action, one may decide to stay at the current dose and collect more data. From this perspective, any dose transition rule can be regarded as a way to interpret statistical evidence in the available DLT data with respect to the key scientific question stated above.

This article proposes to use generalized likelihood ratios (GLRs) to interpret and quantify statistical evidence in making dose transition decisions. The GLR is a measure of statistical evidence for comparing two composite hypotheses according to the generalized law of likelihood \citep{b08,zz13}, which generalizes the law of likelihood for comparing two simple hypotheses \citep{h65,r97,b02}. When applied to a single-dose likelihood in the dose transition problem, the generalized law of likelihood leads to an interval design based on the GLR. Different interval designs can often be calibrated to match or mimic each other. Likewise, the GLR-based interval design can be related to the other interval designs by calculating GLR values at decision boundaries for the other designs. Through these connections, the GLR-based interval design gives a likelihood interpretation to each existing interval design and provides a unified framework for comparing different interval designs in terms of how much evidence they require for escalation and de-escalation. This comparison reveals major differences between different designs with important implications on patient safety, and motivates new configurations of interval designs (based on the GLR) with the potential to improve patient safety.

The GLR-based approach can also be applied to a joint likelihood based on all available DLT data at different dose levels. Such a joint likelihood may be obtained under a nonparametric model (e.g., isotonic regression) or a parsimonious parametric model (e.g., logistic regression) with a smaller number of parameters than the number of doses. For simplicity and ease of presentation, we will focus on the single-dose likelihood initially before considering the joint likelihood and comparing their operating characteristics.

There is increasing awareness that dose-finding studies should consider both toxicity and activity, especially for immune and targeted therapies in oncology with a non-monotone dose-activity relationship \citep{c15,w18,s21,fda23}. This awareness has motivated a new class of designs aiming to find the optimal biological dose \citep{z14,wt15,l16,a18,as18,c18,t18,l20}. The generalized law of likelihood can be applied to accumulating activity data to evaluate statistical evidence about the dose-activity relationship; in fact, that was the original motivation for this research. Designing a dose-finding study that incorporates activity data requires careful considerations for both toxicity and activity assessments, each of which deserves considerable attention. This article, with focus on toxicity, can be viewed a first step in developing a GLR-based design for finding the optimal biological dose \citep{z24}.

The rest of the article is organized as follows. Section \ref{glr} gives a brief review of the generalized law of likelihood, and Section \ref{stat.hyp} formulates statistical hypotheses. In Section \ref{glr.sd}, the GLR is derived on the basis of the single-dose likelihood and used to interpret and compare existing interval designs. Section \ref{glr.iso} defines the joint likelihood and the corresponding GLR. Section \ref{sim} reports simulation studies comparing different designs. The article ends with a discussion in Section \ref{disc}.

%\section{Methods}

\section{Generalized Law of Likelihood}\label{glr}

Suppose data are observed under a parametric model with parameter vector $\bmtheta\in\Theta$. Let $L(\bmtheta)$ be the likelihood for $\bmtheta$ based on the observed data, and let $\bmtheta_1$ and $\bmtheta_2$ be distinct values in $\Theta$. According to the law of likelihood \citep{h65}, the observed data provide statistical evidence supporting the hypothesis $\bmtheta=\bmtheta_1$ over the hypothesis $\bmtheta=\bmtheta_2$ if $L(\bmtheta_1)>L(\bmtheta_2)$, and the likelihood ratio, $L(\bmtheta_1)/L(\bmtheta_2)$, measures the strength of that evidence. From a Bayesian perspective, the likelihood ratio can be interpreted as an invariant Bayes factor for an arbitrary prior distribution of $\bmtheta$. A forceful argument for the law of likelihood is given by Royall \citep{r97}, who also suggests benchmarks for describing the strength of evidence (32 for strong evidence; 8 for moderate evidence). It is worth noting that the likelihood ratio measures statistical evidence in a continuous manner, and the benchmarks are merely to facilitate communication.

%The original law of likelihood assumes a parametric model and is not directly applicable to nonparametric and semiparametric models. For example, if $\theta$ is the mean of a random variable with an unspecified distribution, a parametric likelihood function is unavailable for measuring evidence about $\theta$. This difficulty can be overcome by employing an empirical likelihood function \citep{o88,ql94}, which is readily available for the mean and many other parameters defined using estimating equations, and behaves like a parametric likelihood function for the purpose of measuring statistical evidence \citep{z09}.

The original law of likelihood is restricted to a pair of simple hypotheses, while in practice the hypotheses of interest are often composite in nature, as is the case in dose-finding studies. \citet{b08} and \citet{zz13} independently propose a generalization of the law of likelihood to accommodate composite hypotheses. Let $\Theta_1$ and $\Theta_2$ be subsets of $\Theta$. According to the generalized law of likelihood, the observed data provide statistical evidence supporting the hypothesis $\bmtheta\in\Theta_1$ over the hypothesis $\bmtheta\in\Theta_2$ if $\sup L(\Theta_1):=\sup\{L(\bmtheta):\bmtheta\in\Theta_1\}>\sup L(\Theta_2)$, and the GLR, $\sup L(\Theta_1)/\sup L(\Theta_2)$, measures the strength of that evidence. \citet{zz13} provide an axiomatic development for this generalization, discuss its practical implications, and show that the GLR has reasonable interpretations and properties. With composite hypotheses, the Bayes factor generally depends on the prior distribution of $\bmtheta$, making it impossible to maintain the invariant Bayes factor interpretation of the likelihood ratio. There are also pathological examples in which the (generalized) law of likelihood is of questionable value. Nonetheless, real life examples demonstrate that the GLR measures statistical evidence in a manner consistent with common statistical practice, at least in the context of late phase clinical trials \citep{zz13}.

\section{Statistical Hypotheses}\label{stat.hyp}

Let $p$ denote the true DLT rate of the current dose, and $\phi$ the target DLT rate, i.e., the highest DLT rate that is considered tolerable. To choose an appropriate action (escalate, de-escalate, or stay) for the next step, we need to evaluate the available evidence with respect to two competing hypotheses:
\begin{equation}\label{hyp}
H_1:p\le\phi\qquad\text{versus}\qquad H_2:p>\phi.
\end{equation}
If $H_1$ is known to be true, the correct action is to escalate to the next higher dose. If $H_2$ is known to be true, the correct action is to de-escalate to the next lower dose. %As a philosophical point, if the true value of $p$ is known, there is no need to stay at the current dose because there is nothing more to learn (as far as DLT is concerned).

This formulation of hypotheses differs from previous formulations in the literature on dose-finding studies. Some authors have used simple hypotheses to motivate study designs. For example, the BOIN design with a non-informative prior is essentially a likelihood comparison among three simple hypotheses, and the toxicity assessment in \citet{c18} compares two simple hypotheses using the law of likelihood. These authors demonstrate that simple hypotheses can be useful tools for motivating and modifying study designs to achieve desired operating characteristics. On the other hand, simple hypotheses, especially those chosen on the basis of operating characteristics, may be difficult to interpret as they do not correspond to \lq\lq real" scientific questions. %Some interval designs are based on three interval hypotheses corresponding to three different actions (escalate, stay or de-escalate). The middle interval, commonly known as the equivalence interval, is supposed to contain values of $p$ for which the correct action is to stay. Clinical equivalence may well exist; however, from a philosophical point of view, one might question the existence of an interval (or any value) that supports the decision to stay.  As noted earlier, if the true value of $p$ is known, regardless of the value, there is no point staying at the current dose. From this perspective, the decision to stay results from our uncertainty about $p$ rather than the true value of $p$.

\section{Single-Dose Likelihood}\label{glr.sd}

\begin{figure}
\centering
\includegraphics[width=0.75\textwidth]{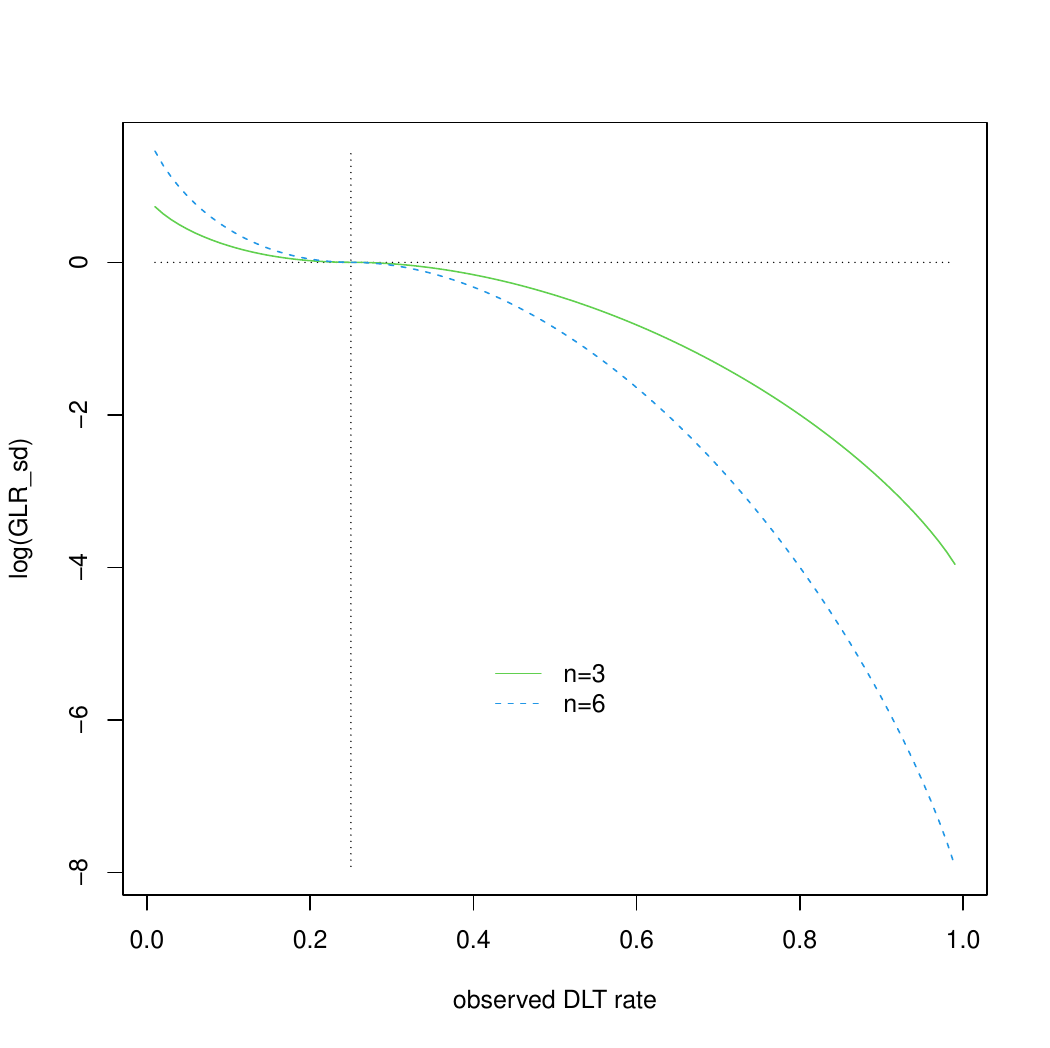}
\caption{$\log(\text{GLR}_{\text{sd}})$ as a function of the observed DLT rate ($\widehat p$) with $\phi=0.25$ and $n=3,6$. The two curves cross each other at $(\widehat p=\phi,\text{GLR}_{\text{sd}}=1)$, as indicated by the dotted vertical and horizontal lines.}
\label{log.glr}
\end{figure}

\noindent Suppose $n$ patients have been treated at the current dose, with DLT events observed in $x$ patients. Assuming the $n$ patients are independent of each other, the likelihood for $p$ is a standard binomial likelihood:
$$
L(p)=p^x(1-p)^{n-x},
$$
which is maximized at $p=x/n=:\widehat p$, the observed DLT rate. The GLR for comparing $H_1$ with $H_2$ is given by $\text{GLR}_{\text{sd}}=\sup_{H_1}L(p)/\sup_{H_2}L(p)$, where
$$
\sup_{H_1}L(p)=\begin{cases}
L(\widehat p)&\text{if }\widehat p\le\phi;\\
L(\phi)&\text{if }\widehat p>\phi,
\end{cases}
\qquad\text{and}\qquad
\sup_{H_2}L(p)=\begin{cases}
L(\phi)&\text{if }\widehat p\le\phi;\\
L(\widehat p)&\text{if }\widehat p>\phi.
\end{cases}
$$
It follows that
\begin{equation}\label{glr.formula}
\text{GLR}_{\text{sd}}=\begin{cases}
\left(\frac{\widehat p}{\phi}\right)^x\left(\frac{1-\widehat p}{1-\phi}\right)^{n-x}
=\left(\frac{\widehat p}{\phi}\right)^{n\widehat p}\left(\frac{1-\widehat p}{1-\phi}\right)^{n(1-\widehat p)}&
\text{if }\widehat p<\phi;\\
1&\text{if }\widehat p=\phi;\\
\left(\frac{\phi}{\widehat p}\right)^x\left(\frac{1-\phi}{1-\widehat p}\right)^{n-x}
=\left(\frac{\phi}{\widehat p}\right)^{n\widehat p}\left(\frac{1-\phi}{1-\widehat p}\right)^{n(1-\widehat p)}
&\text{if }\widehat p>\phi.
\end{cases}
\end{equation}
Although $\widehat p$ is a discrete random variable, $\text{GLR}_{\text{sd}}$ can be regarded as a continuous function of $\widehat p$. Differentiating $\log(\text{GLR}_{\text{sd}})$ with respect to $\widehat p$ shows that $\text{GLR}_{\text{sd}}$ is a decreasing function of $\widehat p$. The rate at which $\log(\text{GLR}_{\text{sd}})$ decreases with $\widehat p$ is proportional to $n$. Figure \ref{log.glr} displays some examples of $\log(\text{GLR}_{\text{sd}})$ as a function of $\widehat p$.

Table \ref{glr.val} shows $\text{GLR}_{\text{sd}}$ values for some common choices of $n$ (3--6) and $\phi$ (0.2, 0.25 and 0.3). In these scenarios, the only way to obtain strong evidence in either direction ($\text{GLR}_{\text{sd}}\ge32$ or $\le1/32$) is when most patients in a cohort have DLT events, in which case $H_2$ is strongly supported over $H_1$. The best possible evidence to support $H_1$ over $H_2$, corresponding to $x=0$, is typically weak ($\text{GLR}_{\text{sd}}<8$) with only one exception ($n=6$, $\phi=0.3$), where it narrowly meets Royal's benchmark for moderate evidence. For intermediate values of $x$, $\text{GLR}_{\text{sd}}$ is typically between 2 and $1/2$. For example, with one out of six patients experiencing DLT events, $\text{GLR}_{\text{sd}}$ ranges between 1.02 and 1.33 as $\phi$ ranges between 0.2 and 0.3. These results indicate that one has to be realistic about how much evidence to expect in a typical dose-finding study. If feasible at all, increasing $n$ beyond 6 is expected to produce better evidence.

The generalized law of likelihood suggests the following dose transition rule based on $\text{GLR}_{\text{sd}}$:
\begin{itemize}
\item Escalate if $\text{GLR}_{\text{sd}}\ge k_1$;
\item De-escalate if $\text{GLR}_{\text{sd}}\le 1/k_2$;
\item Stay if $1/k_2<\text{GLR}_{\text{sd}}<k_1$.
\end{itemize}
The cut-points $k_1$ and $k_2$ represent the levels of evidence required for escalation and de-escalation; they need not be equal and may be chosen on the basis of practical considerations. The sample size of a dose-finding study is typically small (3--9 per dose), which suggests that $k_1$ and $k_2$ cannot be too large. If incorrect escalation is considered more severe than incorrect de-escalation, one may choose $k_1>k_2$, which indicates that more evidence is required for escalation than for de-escalation. Because $\text{GLR}_{\text{sd}}$ is a decreasing function of $\widehat p$, the above transition rule leads to an interval design where small values of $\widehat p$ lead to escalation, large values to de-escalation, and intermediate values to a \lq\lq stay" decision. Different interval designs can often be calibrated to match or mimic each other by choosing appropriate parameter values. Likewise, the interval design based on $\text{GLR}_{\text{sd}}$ can be related to the other interval designs by calculating values of $\text{GLR}_{\text{sd}}$ at decision boundaries for the other designs. In this manner, the interval design based on $\text{GLR}_{\text{sd}}$ gives a likelihood interpretation to each existing interval design and provides a unified framework in which different interval designs with the same target DLT rate can be compared directly in terms of how much evidence is required for dose escalation and de-escalation.

For the same choices of $(n,\phi)$ considered in Table \ref{glr.val}, Table \ref{k.val} reports the \lq\lq effective" values of $(k_1,k_2)$ of the 3+3 and various interval designs, regarded as interval designs based on $\text{GLR}_{\text{sd}}$ with the same target DLT rate and the same decision boundaries in $\widehat p$. The 3+3 design is not really an interval design; it does not have a clearly stated target DLT rate or precisely defined decision boundaries in $\widehat p$. To ease the comparison, for each given value of $\phi$, we use the results in Table \ref{glr.val} to determine the ranges of possible values of $(k_1,k_2)$ for interval designs based on $\text{GLR}_{\text{sd}}$ that make the same decisions as the 3+3 design for $n=3$ and $n=6$. The BOIN, TEQR and i3+3 designs have well-defined decision boundaries in $\widehat p$, which are substituted into equation \eqref{glr.formula} to compute $(k_1,k_2)$. By replacing $x$ with $n\widehat p$ in the posterior distribution of $p$, the mTPI design can be extended easily to accommodate a continuous $\widehat p$ and thus can be treated in the same manner. For the BOIN design, we set $\phi_1=0.6\phi$ and $\phi_2=1.4\phi$ as recommended by \citet{ly15}. For designs that require an equivalence interval, it is chosen to be $(\phi-0.05,\phi+0.05)$, consistent with common practice. A uniform prior is used in the mTPI design.

In Table \ref{k.val}, all values of $(k_1,k_2)$ are below 2, and most are below 1.5, representing a much lower level of evidence than what is commonly expected in phase III trials ($k=6.82$) \citep{zz13}. The lower requirement may be appropriate for the objective of the study (dose finding as opposed to hypothesis testing) and the typically small sample size. The likelihood interpretation of the 3+3 design depends heavily on the target DLT rate ($\phi$). With $\phi=0.2$, the design requires more evidence for de-escalation than for escalation. With $\phi=0.3$, it does the opposite. With $\phi=0.25$, it requires similar amounts of evidence for escalation and de-escalation. The BOIN and TEQR designs have similar values of $(k_1,k_2)$, which are quite close to 1, and both designs require similar amounts of evidence for escalation and de-escalation, regardless of $\phi$. The mTPI and i3+3 designs are similar in that both require much more evidence for de-escalation than for escalation. For escalation, the requirement of the i3+3 design is lower than that of the mTPI design and similar to those of the BOIN and TEQR designs.

By setting a higher bar for de-escalation than for escalation, the mTPI and i3+3 designs tend to be aggressive in moving toward higher doses. This has a potential impact on patient safety, which will be examined in a simulation study. It seems difficult to imagine a general rationale for requiring more evidence for de-escalation than for escalation. In fact, from a patient safety perspective, there is a good rationale for requiring more evidence for escalation than for de-escalation. DLT events are serious events, and it would be advisable to use some caution in dose escalation. No design in Table \ref{k.val} consistently requires more evidence for escalation than for de-escalation. Although some designs can be modified to produce that effect (e.g., by specifying an asymmetric equivalence interval), such a modification would be ad hoc and unintuitive. In the GLR-based interval design, one can achieve that goal in a simple and transparent manner by setting $k_1>k_2$. This possibility will be explored in a simulation study.

\section{Joint Likelihood}\label{glr.iso}

The single-dose likelihood used in $\text{GLR}_{\text{sd}}$ is based on observed DLT data at one dose only. A joint likelihood can be used to borrow information across doses under a suitable model. Suppose we have collected DLT data for the first $d$ doses: $\{(n_i,x_i):i=1,\dots,d\}$, where $n_i$ is the number of patients treated at dose $i$ and $x_i$ is the number of patients who experienced DLT events. Write $\bmp=(p_1,\dots,p_d)'$, where $p_i$ is the true DLT rate of dose $i\in\{1,\dots,d\}=:[d]$. The joint likelihood for $\bmp$ is simply a product of dose-specific binomial likelihoods:
$$
L(\bmp)=\prod_{i=1}^dp_i^{x_i}(1-p_i)^{n_i-x_i}.
$$
For the current dose $c\in[d]$, the two competing hypotheses are $H_{1}:p_c\le\phi$ versus $H_{2}:p_c>\phi$. Without any constraints on $\bmp$, the suprema of $L(\bmp)$ over $H_1$ and $H_2$ are proportional to those based on the single-dose likelihood for $p_c$, leading to the same GLR for comparing $H_1$ with $H_2$. However, the GLR may differ if $\bmp$ is constrained.

Let us first consider the isotonic regression model where $\bmp$ is confined to $\mathcal P=\{\bmp:0\le p_1\le\cdots\le p_d\le1\}$. Maximizing $L(\bmp)$ over $\bmp\in\mathcal P$ yields a constrained maximum likelihood estimate of $\bmp$, say $\widetilde\bmp=(\widetilde p_1,\dots,\widetilde p_d)'$, which can be used to estimate the MTD at the end of the study. Specifically, the estimated MTD is $\max\{i\in[d]:\widetilde p_i\le\phi\}$, which is set to 0 if there is no qualifying $i$. 
The GLR for comparing $H_{1}$ with $H_{2}$ is $\text{GLR}_{\text{iso}}=\sup_{H_{1}}L(\bmp)/\sup_{H_{2}}L(\bmp)$, where
\begin{equation*}\begin{aligned}
\sup_{H_{1}}L(\bmp)&=\max\{L(\bmp):\bmp\in\mathcal P,p_c\le\phi\},\\
\sup_{H_{2}}L(\bmp)&=\max\{L(\bmp):\bmp\in\mathcal P,p_c\ge\phi\}.
\end{aligned}\end{equation*}
The maximizations involved here and in finding $\widetilde\bmp$ are convex optimization problems with linear inequality constraints, which can be solved using an adaptive barrier algorithm (available in R as \texttt{constrOptim()}). Once computed, $\text{GLR}_{\text{iso}}$ can be compared to $(k_1,k_2)$ to make a dose transition decision, in the same manner as $\text{GLR}_{\text{sd}}$ is used. As an illustrative example, suppose $\phi=0.25$, $c=d=3$, $(n_1,n_2,n_3)=(3,6,3)$ and $(x_1,x_2,x_3)=(0,1,0)$; then $\widetilde\bmp\approx(0,0.11,0.11)'$ and $\text{GLR}_{\text{iso}}\approx1.53$, less than $\text{GLR}_{\text{sd}}\approx2.37$ (given in Table \ref{glr.val}). The reduction in the GLR for dose 3 acknowledges the fact that a DLT event has been observed at dose 2.

One could also use a parsimonious parametric likelihood if there is enough prior information to support it. Suppose $\bmp$ is modeled as $\bmp(\bmtheta)$, where $\bmtheta\in\Theta$ is a parameter vector of lower dimension than $\bmp$. The joint likelihood based on this model is
$$
L(\bmtheta)=\prod_{i=1}^dp_i(\bmtheta)^{x_i}(1-p_i(\bmtheta))^{n_i-x_i}.
$$
Let $\widehat\bmtheta$ denote the maximum likelihood estimate of $\bmtheta$, which maximizes the above likelihood over $\bmtheta\in\Theta$; then $\bmp$ is estimated by $\bmp(\widehat\bmtheta)$ and the MTD by $\max\{i\in[d]:p_i(\widehat\bmtheta)\le\phi\}$, which defaults to 0 if there is no qualifying $i$. For comparing $H_1$ with $H_2$ at the current dose, the model-based GLR is $\sup_{H_{1}}L(\bmtheta)/\sup_{H_{2}}L(\bmtheta)$, where
\begin{equation*}\begin{aligned}
\sup_{H_{1}}L(\bmtheta)&=\sup\{L(\bmtheta):\bmtheta\in\Theta,p_c(\bmtheta)\le\phi\},\\
\sup_{H_{2}}L(\bmtheta)&=\sup\{L(\bmtheta):\bmtheta\in\Theta,p_c(\bmtheta)>\phi\}.
\end{aligned}\end{equation*}
In statistical estimation, a model-based estimator is typically more efficient than a nonparametric one when the underlying model is correct, and potentially biased when the model is incorrect. The model-based GLR is subject to a similar trade-off, which will be examined in a simulation study.

\section{Simulation Studies}\label{sim}

\subsection{Single-Dose GLR vs Other Interval Designs}\label{sim1}

This first simulation study evaluates the proposed design based on $\text{GLR}_{\text{sd}}$ with $k_1>k_2$ in comparison with the BOIN, TEQR, mTPI and i3+3 designs. The $\text{GLR}_{\text{sd}}$-based design is implemented with cut-points $k_1=1.5$ and $k_2=1.05$ or 1.1, chosen to represent the \lq\lq opposite" of the mTPI and i3+3 designs in the sense that more evidence is required for escalation than for de-escalation. The BOIN design is implemented with the default setting: $\phi_1=0.6\phi$ and $\phi_2=1.4\phi$. The TEQR, mTPI and i3+3 designs are implemented with the same equivalence interval: $(\phi-0.05,\phi+0.05)$. The uniform prior is used in the mTPI design. All designs start at the lowest dose and include an overdose control mechanism based on posterior probabilities. Specifically, a dose (and all higher doses) will be eliminated if $\pr(p>\phi|n,x)>0.95$ under a uniform prior for $p$, as suggested by \citet{ly15}. For the proposed design, an alternative condition for dose elimination is $\text{GLR}_{\text{sd}}\le1/3.87$, equivalent to a significant likelihood ratio test at $\alpha=0.05$ with $H_1:p\le\phi$ as the null hypothesis. This alternative condition appears to have little impact on operating characteristics, and the corresponding results are not reported here. For all designs, MTD estimation is performed by applying constrained maximum likelihood estimation under the isotonic regression model (see section \ref{glr.iso}) to all available data at the end of the study.

\begin{figure}
\centering
\includegraphics[width=0.75\textwidth]{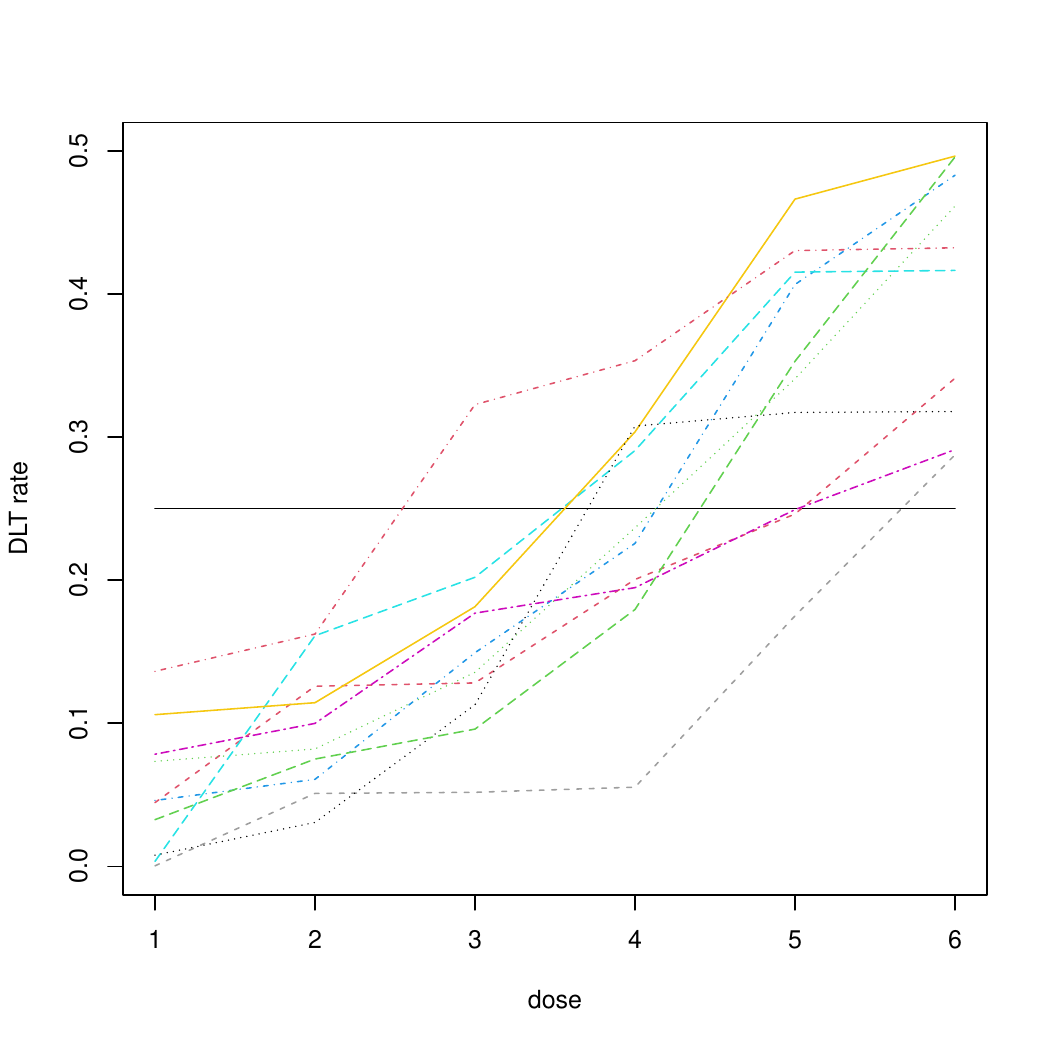}
\caption{10 sets of true DLT rates at $D=6$ doses, generated randomly as in the simulation study of Section \ref{sim1} with $\gamma=2$ and $\phi=0.25$ (indicated by the solid horizontal line).}
\label{random.pp}
\end{figure}

\begin{figure}
\centering
\includegraphics[width=0.85\textwidth]{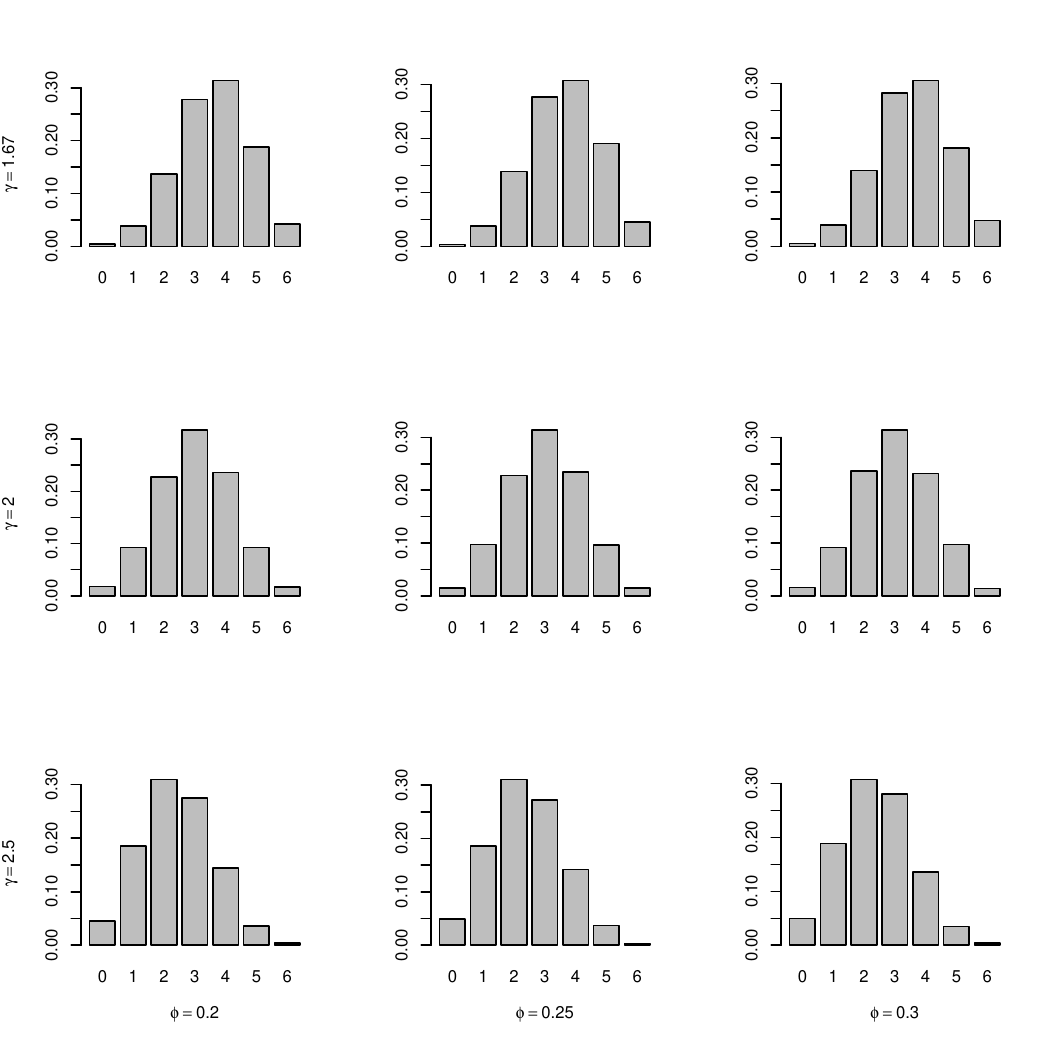}
\caption{Empirical distributions (across $10^4$ simulations) of the true MTD among $D=6$ candidate doses and 0 (for \lq\lq no MTD") for some values of $(\gamma,\phi)$ in the simulation study of Section \ref{sim1}.}
\label{pct.mtd}
\end{figure}

In this simulation study, the cohort size is fixed at 3, the total number of doses is $D\in\{4,6,8\}$, and for each value of $D$ the maximum number of cohorts is $M=2D$ (i.e., 2 cohorts per dose on average). The target DLT rate is $\phi\in\{0.2,0.25,0.3\}$. Given $\phi$ and $D$, the true DLT rates, $\bmp=(p_1,\dots,p_D)'$, are ordered values in a random sample of size $D$ from the uniform distribution on $(0,\gamma\phi)$, where $\gamma\in\{5/3,2,5/2\}$. The vector $\bmp$ is generated anew for each trial. Figure \ref{random.pp} shows a random sample of 10 realizations of $\bmp$ with $D=6$, $\phi=0.25$ and $\gamma=2$. Given $\phi$ and $\bmp$, the true MTD is defined as $\max\{i\in[D]:p_i\le\phi\}$, with the understanding that the maximum of an empty set is 0. Figure \ref{pct.mtd} shows the distribution of the MTD (based on $10^4$ simulations) for $D=6$, $\gamma\in\{5/3,2,5/2\}$ and $\phi\in\{0.2,0.25,0.3\}$. As shown in Figure \ref{pct.mtd}, the shape of the MTD distribution is mainly driven by the value of $\gamma$. In each scenario defined by $(\gamma,D,\phi)$, a total of $10^4$ trials are simulated for each design.

The different designs are compared using three metrics: the percentage of trials that correctly find the MTD (which may be 0), the percentage of patients who are over-treated (i.e., treated at a dose above the true MTD), and the average number of patients treated, which may be less than $N_{\max}=3M=6D$ due to the possibility of early stopping. The three metrics represent, respectively, the accuracy of MTD estimation, the frequency of over-treatment, and the cost of the study.

Using the three metrics defined above, the simulation results are summarized in Tables \ref{sim.rst.1}--\ref{sim.rst.3} for $\gamma=5/3$, 2 and 5/2, respectively. In general, MTD estimation accuracy increases with $\phi$ and decreases with $D$. In each scenario, the average number of patients treated is typically no less than $N_{\max}-2$ and varies little across designs. The different designs do differ substantially in the frequency of over-treatment. Typically, the mTPI and i3+3 designs over-treat more frequently than the BOIN and TEQR designs, which in turn over-treat more frequently than the proposed design based on $\text{GLR}_{\text{sd}}$. In terms of MTD estimation, the interval designs generally perform similarly to each other, except that when $\gamma=5/3$ the proposed design may under-perform the other designs slightly in some scenarios. This is not surprising because $\gamma=5/3$ skews the distribution of the true MTD toward higher doses (see Figure \ref{pct.mtd}) and thus favors designs that escalate more aggressively. In this case, the proposed design may be associated with a trade-off between a possible slight decrease in MTD estimation accuracy and a predictable decrease in over-treatment. For higher values of $\gamma$, the proposed design compare favorably with the other interval designs, with similar accuracy of MTD estimation and lower frequency of over-treatment. Within the proposed design, there is an expected trade-off between $k_2=1.05$ and $k_2=1.1$: the smaller value is more effective for reducing over-treatment but may result in a slight decrease in MTD estimation accuracy.

\subsection{Single-Dose vs Joint Likelihood Designs}\label{sim2}

We now compare the $\text{GLR}_{\text{sd}}$-based design with various designs based on the joint likelihood for all observed data (across dose levels). Specifically, the comparison includes $\text{GLR}_{\text{iso}}$ (based on isotonic regression), $\text{GLR}_{\text{pow}}$ (based on a power model used in the CRM design), and $\text{GLR}_{\text{lgx}}$ (based on a logistic regression model). The power model is specified as $p_i(\theta)=p_{0i}^{\theta}$, where $\theta\in(0,\infty)$ is an unknown parameter and $p_{0i}=2i\phi/(D+1)$ is the prior mean of $p_i$ according to the data generation mechanism in Section \ref{sim1} with $\gamma=2$. The logistic regression model is specified as $p_i(\alpha,\beta)=\expit(\alpha+\beta z_i^*)$, where $\expit(u)=\{1+\exp(-u)\}^{-1}$, $\alpha\in(-\infty,\infty)$ and $\beta\in[0,\infty)$ are unknown parameters, and $z_i^*$ is a linear transformation of the logarithm of the original dose value $z_i$:
$$
z_i^*=\frac{\log(z_i)-\log(z_1)}{\log(z_D)-\log(z_1)}=\frac{\log(z_i/z_1)}{\log(z_D/z_1)}.
$$
The original dose values are assumed to follow geometric progression: $z_i=z_1\rho^{i-1}$ for some $z_1>0$ and $\rho>1$. Regardless of the values of $(z_1,\rho)$, we have $z_i^*=(i-1)/(D-1)\in[0,1]$, $i=1,\dots,D$. Thus, working with $z_i^*$ instead of $\log(z_i)$ ensures that the regression parameters $(\alpha,\beta)$ have clear interpretations invariant to $(z_1,\rho)$. All GLR-based designs are implemented with $k_1=1.5$ and $k_2=1.05$ as well as the same overdose control mechanism described in Section \ref{sim1}. \textcolor{blue}{Appendix A in the Supplementary Material explains why this comparison does not include the CRM and BLRM designs.}

These \textcolor{blue}{GLR-based} designs are compared in three different scenarios in which the power and logistic models may be correct or incorrect. In Scenario 1, $\bmp$ is generated from the power model with $\log(\theta)\sim N(0,1.34)$. In Scenario 2, $\bmp$ is generated from the logistic model with $\alpha\sim N(\logit(\phi)-1/2,9/16)$ and $\beta\sim\max\{0,N(1,1/4)\}$, independently of each other. The prior mean of $\alpha$ is chosen such that the target DLT rate $\phi$ is attained at $z^*=1/2$ for a typical dose-toxicity curve with $(\alpha,\beta)$ equal to their median values. The truncation of $\beta$ at 0 ensures $\beta\ge0$, and the case $\beta=0$ corresponds to a dose-independent toxicity profile, which occurs with a small probability ($\approx0.025$). In Scenario 3, $\bmp$ is generated as in Section \ref{sim1} with $\gamma=2$, which is inconsistent with both models. In each scenario, $\bmp$ is generated anew for each of $10^4$ simulated trials. Except for the generation of $\bmp$, the three scenarios are identical to each other in all other aspects, including $D\in\{4,6,8\}$ and $\phi\in\{0.2,0.25,0.3\}$.

The results of this simulation study are summarized in Table \ref{sim.rst.4} using the same three metrics reported in Tables \ref{sim.rst.1}--\ref{sim.rst.3}. The nonparametric designs based on $\text{GLR}_{\text{sd}}$ and $\text{GLR}_{\text{iso}}$ preform similarly in most cases; any differences are small and inconsistent. It is not clear that using the join likelihood based on the isotonic regression model instead of the single-dose likelihood leads to improved performance. The parametric designs based on $\text{GLR}_{\text{pow}}$ and $\text{GLR}_{\text{lgx}}$ do perform differently than the nonparametric designs. When the underlying power model is correct (Scenario 1), the $\text{GLR}_{\text{pow}}$-based design performs best in terms of MTD estimation accuracy while producing more over-treatment than the other designs. When the underlying logistic model is correct (Scenario 2), the $\text{GLR}_{\text{lgx}}$-based design outperforms the nonparametric designs in MTD estimation accuracy. Interestingly, despite model mis-specification, the $\text{GLR}_{\text{pow}}$-based design appears quite competitive in Scenario 2 and often outperforms the $\text{GLR}_{\text{lgx}}$-based design at lower values of $\phi$ and/or $D$. One possible explanation is that the mis-specification of the power model is not severe enough in this scenario to outweigh the benefit of parsimony. Here again, the competitive performance (in MTD estimation) of the $\text{GLR}_{\text{pow}}$-based design comes at the expense of more frequent over-treatment. In Scenario 3, where both parametric models are mis-specified, both parametric designs underperform the nonparametric designs in terms of MTD estimation accuracy. In this scenario, the $\text{GLR}_{\text{pow}}$-based design consistently underperforms the $\text{GLR}_{\text{lgx}}$-based design, suggesting that the power model is more severely mis-specified than the logistic model. It is worth noting that the $\text{GLR}_{\text{pow}}$-based design continues to produce more over-treatment in Scenario 3 despite lower performance in MTD estimation.

\section{Discussion}\label{disc}

This article provides a likelihood perspective on dose transition rules in oncologic dose-finding studies that aim to find the MTD. In this perspective, the key scientific question is whether the true DLT rate of the current dose exceeds the target DLT rate, and the statistical question is how to evaluate the statistical evidence in the available DLT data with respect to that scientific question. According to the generalized law of likelihood, the GLR can be used as a measure of statistical evidence and as the basis for making dose transition decisions. This leads to a GLR-based dose transition rule that can be used to interpret and compare existing interval designs and construct new ones. A GLR-based comparison of commonly used interval designs indicates that some designs require more evidence for de-escalation than for escalation, which raises concerns about patient safety. Simulation results confirm that those designs frequently produce higher proportions of over-treated patients. These observations motivate the consideration of GLR-based designs that require more evidence for escalation than for de-escalation, which are shown in a simulation study to reduce the proportion of over-treated patients while achieving a similar level of MTD estimation accuracy.

The GLR approach can be applied to a single-dose likelihood or a joint likelihood based on a nonparametric (e.g., isotonic regression) model or a parametric model. Simulation results demonstrate that a parametric model-based GLR can improve MTD estimation (possibly at the expense of increased over-treatment) when the underlying model is correct. Such a parametric design may be preferable (over nonparametric ones) if the investigator is confident about the model and comfortable with its operating characteristics. Without enough prior information to support a parsimonious model, one might choose a design based on a nonparametric GLR, which tends to be more robust.

Compared to the single-dose likelihood, the joint likelihood based on the isotonic regression model does not appear to enhance the performance of the GLR-based design. This is not surprising because the monotonicity assumption is fairly weak and confers limited ability to borrow information across doses. From a practical point of view, $\text{GLR}_{\text{sd}}$ is easy to compute and the resulting dose transition rule can be tabulated before starting a study. Computing $\text{GLR}_{\text{iso}}$ is more difficult and requires special software, and the computation has to be conducted in real time if one chooses to use this approach in a study. These observations, together with our simulation results, support the use of $\text{GLR}_{\text{sd}}$ instead of $\text{GLR}_{\text{iso}}$ if a nonparametric GLR is to be used to design a dose-finding study.

%While this article is focused on toxicity assessment, the generalized law of likelihood can also be used to address research questions related to drug activity. There is an ongoing effort to formulate statistical hypotheses on drug activity, derive and compute the corresponding generalized likelihood ratios, and develop dose transition rules that take both toxicity and activity into account.

\pagebreak

\pagebreak

\pagebreak

%\pagebreak
%\begin{landscape}
\renewcommand{\baselinestretch}{1.2}
\begin{table}[htbp]
%{\small
\caption{Values of $\text{GLR}_{\text{sd}}$ for comparing $H_1:p\le\phi$ with $H_2:p>\phi$ with common choices of $n$ and $\phi$. $\text{GLR}_{\text{sd}}$ values greater than 1 support $H_1$ over $H_2$. $\text{GLR}_{\text{sd}}$ values less than 1, which support $H_2$ over $H_1$, are expressed as reciprocals to indicate the strength of support.}\label{glr.val}
\newcolumntype{d}{D{.}{.}{2}}
\begin{center}
\begin{tabular}{ccccc}
\hline
\hline
\multicolumn{1}{c}{$n$}&\multicolumn{1}{c}{$x$}&\multicolumn{3}{c}{$\phi$}\\
\cline{3-5}
\multicolumn{2}{c}{}&\multicolumn{1}{c}{0.2}&\multicolumn{1}{c}{0.25}&\multicolumn{1}{c}{0.3}\\
\hline
3&0&1.95&2.37&2.92\\
&1&1/1.16&1/1.05&1/1.01\\
&2&1/4.63&1/3.16&1/2.35\\
&3&$<1/100$&1/64.0&1/37.0\\
\hline
4&0&2.44&3.16&4.16\\
&1&1/1.03&1.00&1.02\\
&2&1/2.44&1/1.78&1/1.42\\
&3&1/16.5&1/9.0&1/5.58\\
&4&$<1/100$&$<1/100$&$<1/100$\\
\hline
5&0&3.05&4.21&5.95\\
&1&1.00&1.04&1.14\\
&2&1/1.69&1/1.31&1/1.12\\
&3&1/6.75&1/3.93&1/2.61\\
&4&1/64.0&1/28.0&1/14.4\\
&5&$<1/100$&$<1/100$&$<1/100$\\
\hline
6&0&3.81&5.62&8.50\\
&1&1.02&1.13&1.33\\
&2&1/1.34&1/1.11&1/1.02\\
&3&1/3.81&1/2.37&1/1.69\\
&4&1/21.4&1/9.99&1/5.53\\
&5&$<1/100$&1/91.4&1/39.4\\
&6&$<1/100$&$<1/100$&$<1/100$\\
\hline
\end{tabular}
\end{center}
%}
\end{table}
%\end{landscape}

%\pagebreak
%\begin{landscape}
\renewcommand{\baselinestretch}{1.2}
\begin{table}[htbp]
%{\small
\caption{Effective values of $(k_1,k_2)$ for the 3+3 and various interval designs, regarded as interval designs based on $\text{GLR}_{\text{sd}}$ with the same target DLT rate ($\phi$) and the same decision boundaries in $\widehat p$ (see section \ref{glr.sd} for details).}\label{k.val}
\newcolumntype{d}{D{.}{.}{2}}
\begin{center}
\begin{tabular}{cccccccccc}
\hline
\hline
\multicolumn{1}{c}{Design}&\multicolumn{1}{c}{$n$}&\multicolumn{2}{c}{$\phi=0.2$}&\multicolumn{1}{c}{}&\multicolumn{2}{c}{$\phi=0.25$}&\multicolumn{1}{c}{}&\multicolumn{2}{c}{$\phi=0.3$}\\
\cline{3-4}\cline{6-7}\cline{9-10}
\multicolumn{2}{c}{}&\multicolumn{1}{c}{$k_1$}&\multicolumn{1}{c}{$k_2$}&\multicolumn{1}{c}{}&\multicolumn{1}{c}{$k_1$}&\multicolumn{1}{c}{$k_2$}&\multicolumn{1}{c}{}&\multicolumn{1}{c}{$k_1$}&\multicolumn{1}{c}{$k_2$}\\
\hline
3+3&&1.00--1.02&1.16--1.34&&1.00--1.13&1.05--1.11&&1.00--1.33&1.01--1.02\\
\hline
BOIN&3&1.02&1.01&&1.02&1.02&&1.03&1.02\\
&4&1.02&1.02&&1.03&1.02&&1.04&1.03\\
&5&1.03&1.02&&1.04&1.03&&1.05&1.04\\
&6&1.04&1.03&&1.05&1.04&&1.06&1.05\\
\hline
TEQR&3&1.03&1.02&&1.02&1.02&&1.02&1.02\\
&4&1.03&1.03&&1.03&1.03&&1.02&1.02\\
&5&1.04&1.04&&1.04&1.03&&1.03&1.03\\
&6&1.05&1.05&&1.04&1.04&&1.04&1.04\\
\hline
mTPI&3&1.10&1.54&&1.13&1.47&&1.15&1.42\\
&4&1.13&1.68&&1.16&1.61&&1.20&1.54\\
&5&1.16&1.82&&1.20&1.73&&1.24&1.65\\
&6&1.19&1.95&&1.24&1.84&&1.28&1.75\\
\hline
i3+3&3&1.03&1.83&&1.02&1.73&&1.02&1.67\\
&4&1.03&1.52&&1.03&1.46&&1.02&1.42\\
&5&1.04&1.36&&1.04&1.31&&1.03&1.28\\
&6&1.05&1.25&&1.04&1.22&&1.04&1.20\\
\hline
\end{tabular}
\end{center}
%}
BOIN: Bayesian optimal interval; TEQR: toxicity equivalence range; mTPI: modified toxicity probability interval.
\end{table}
%\end{landscape}

\renewcommand{\baselinestretch}{1.2}
\begin{table}[htbp]
{\small
\caption{Results of the first simulation study with $\gamma=5/3$: comparison of existing interval designs with the proposed design based on $\text{GLR}_{\text{sd}}$ with $k_1=1.5$ and $k_2=1.05$ or 1.1, with respect to the accuracy of MTD estimation, the frequency of over-treatment, and the cost of the study (see section \ref{sim1} for details).}\label{sim.rst.1}
\newcolumntype{d}{D{.}{.}{1}}
\begin{center}
\begin{tabular}{cccccccccccccc}
\hline
\hline
\multicolumn{1}{c}{$D$}&\multicolumn{1}{c}{Design}&\multicolumn{1}{c}{$k_2$}&\multicolumn{3}{c}{$\phi=0.2$}&\multicolumn{1}{c}{}&\multicolumn{3}{c}{$\phi=0.25$}&\multicolumn{1}{c}{}&\multicolumn{3}{c}{$\phi=0.3$}\\
\cline{4-6}\cline{8-10}\cline{12-14}
\multicolumn{3}{c}{}&\multicolumn{1}{c}{\%MTD}&\multicolumn{1}{c}{\%OT}&\multicolumn{1}{c}{$N_{\text{ave}}$}&\multicolumn{1}{c}{}&\multicolumn{1}{c}{\%MTD}&\multicolumn{1}{c}{\%OT}&\multicolumn{1}{c}{$N_{\text{ave}}$}&\multicolumn{1}{c}{}&\multicolumn{1}{c}{\%MTD}&\multicolumn{1}{c}{\%OT}&\multicolumn{1}{c}{$N_{\text{ave}}$}\\
\hline
4&BOIN&&41.9&27.4&23.4&&46.5&28.1&23.8&&50.2&31.6&23.7\\
&TEQR&&42.9&27.5&23.4&&46.0&28.2&23.8&&50.2&31.7&23.7\\
&mTPI&&43.3&34.1&23.4&&47.4&31.0&23.8&&50.9&31.7&23.7\\
&i3+3&&43.4&31.4&23.4&&48.4&34.8&23.9&&50.5&32.1&23.7\\
&$\text{GLR}_{\text{sd}}$&1.05&41.5&25.1&23.4&&43.7&23.1&23.8&&49.4&25.3&23.8\\
&$\text{GLR}_{\text{sd}}$&1.1&42.8&25.3&23.4&&46.3&27.0&23.8&&50.0&25.5&23.7\\
\hline
6&BOIN&&34.3&23.1&35.5&&37.2&23.6&35.9&&42.3&27.6&35.8\\
&TEQR&&35.2&23.2&35.5&&38.2&23.3&35.9&&42.6&27.6&35.8\\
&mTPI&&34.7&29.5&35.5&&37.9&25.1&35.9&&41.9&26.5&35.8\\
&i3+3&&34.7&27.2&35.5&&39.3&30.6&35.9&&42.1&27.6&35.9\\
&$\text{GLR}_{\text{sd}}$&1.05&34.3&19.6&35.5&&37.0&17.4&35.9&&41.1&19.8&35.8\\
&$\text{GLR}_{\text{sd}}$&1.1&34.5&20.2&35.6&&37.3&21.3&35.9&&41.2&19.8&35.8\\
\hline
8&BOIN&&27.9&19.1&47.6&&32.4&20.1&48.0&&35.6&23.6&47.9\\
&TEQR&&28.6&20.0&47.6&&32.2&20.5&47.9&&36.2&24.4&47.9\\
&mTPI&&29.0&25.6&47.6&&32.7&21.3&47.9&&35.0&21.9&47.9\\
&i3+3&&29.3&23.8&47.6&&33.5&27.2&47.9&&36.4&24.2&47.9\\
&$\text{GLR}_{\text{sd}}$&1.05&27.0&16.1&47.6&&29.4&14.4&47.9&&34.8&16.3&47.9\\
&$\text{GLR}_{\text{sd}}$&1.1&27.5&16.6&47.6&&31.8&17.4&47.9&&35.3&16.3&47.9\\
\hline
\end{tabular}
\end{center}
}
\%MTD: percentage of trials that correctly find the MTD; \%OT: percentage of patients who are over-treated (i.e., treated at a dose above the true MTD); $N_{\text{ave}}$: average number of patients treated; BOIN: Bayesian optimal interval; TEQR: toxicity equivalence range; mTPI: modified toxicity probability interval.
\end{table}

\renewcommand{\baselinestretch}{1.2}
\begin{table}[htbp]
{\small
\caption{Results of the first simulation study with $\gamma=2$: comparison of existing interval designs with the proposed design based on $\text{GLR}_{\text{sd}}$ with $k_1=1.5$ and $k_2=1.05$ or 1.1, with respect to the accuracy of MTD estimation, the frequency of over-treatment, and the cost of the study (see section \ref{sim1} for details).}\label{sim.rst.2}
\newcolumntype{d}{D{.}{.}{1}}
\begin{center}
\begin{tabular}{cccccccccccccc}
\hline
\hline
\multicolumn{1}{c}{$D$}&\multicolumn{1}{c}{Design}&\multicolumn{1}{c}{$k_2$}&\multicolumn{3}{c}{$\phi=0.2$}&\multicolumn{1}{c}{}&\multicolumn{3}{c}{$\phi=0.25$}&\multicolumn{1}{c}{}&\multicolumn{3}{c}{$\phi=0.3$}\\
\cline{4-6}\cline{8-10}\cline{12-14}
\multicolumn{3}{c}{}&\multicolumn{1}{c}{\%MTD}&\multicolumn{1}{c}{\%OT}&\multicolumn{1}{c}{$N_{\text{ave}}$}&\multicolumn{1}{c}{}&\multicolumn{1}{c}{\%MTD}&\multicolumn{1}{c}{\%OT}&\multicolumn{1}{c}{$N_{\text{ave}}$}&\multicolumn{1}{c}{}&\multicolumn{1}{c}{\%MTD}&\multicolumn{1}{c}{\%OT}&\multicolumn{1}{c}{$N_{\text{ave}}$}\\
\hline
4&BOIN&&46.8&33.8&23.2&&51.5&34.3&23.7&&55.0&37.2&23.5\\
&TEQR&&48.8&34.3&23.2&&51.6&34.2&23.7&&56.0&37.3&23.5\\
&mTPI&&47.5&41.3&23.1&&53.2&37.3&23.7&&55.5&37.5&23.5\\
&i3+3&&47.2&38.5&23.2&&52.5&41.2&23.7&&55.1&37.1&23.5\\
&$\text{GLR}_{\text{sd}}$&1.05&46.3&31.0&23.1&&51.3&29.0&23.7&&55.4&30.7&23.5\\
&$\text{GLR}_{\text{sd}}$&1.1&48.2&31.2&23.1&&51.6&32.9&23.7&&56.4&31.1&23.5\\
\hline
6&BOIN&&38.7&28.1&35.3&&44.1&28.9&35.8&&48.1&32.2&35.7\\
&TEQR&&40.0&28.6&35.4&&44.8&28.7&35.7&&47.1&33.1&35.6\\
&mTPI&&39.2&34.9&35.2&&44.3&30.5&35.8&&48.1&32.3&35.6\\
&i3+3&&39.6&32.5&35.3&&43.8&35.6&35.8&&47.4&32.3&35.6\\
&$\text{GLR}_{\text{sd}}$&1.05&39.0&24.3&35.2&&42.2&22.5&35.8&&47.3&24.5&35.6\\
&$\text{GLR}_{\text{sd}}$&1.1&39.1&25.1&35.3&&44.0&26.4&35.8&&47.8&24.8&35.7\\
\hline
8&BOIN&&34.1&24.5&47.4&&38.2&25.4&47.8&&43.3&27.9&47.8\\
&TEQR&&34.2&24.6&47.3&&38.7&24.9&47.9&&43.6&29.3&47.8\\
&mTPI&&34.2&30.5&47.4&&38.0&26.8&47.9&&43.1&27.3&47.8\\
&i3+3&&33.9&28.6&47.4&&39.6&32.0&47.9&&43.0&28.9&47.8\\
&$\text{GLR}_{\text{sd}}$&1.05&33.5&20.8&47.4&&36.8&19.0&47.8&&40.8&20.9&47.7\\
&$\text{GLR}_{\text{sd}}$&1.1&34.2&21.4&47.4&&38.4&22.5&47.9&&41.4&21.3&47.8\\
\hline
\end{tabular}
\end{center}
}
\%MTD: percentage of trials that correctly find the MTD; \%OT: percentage of patients who are over-treated (i.e., treated at a dose above the true MTD); $N_{\text{ave}}$: average number of patients treated; BOIN: Bayesian optimal interval; TEQR: toxicity equivalence range; mTPI: modified toxicity probability interval.
\end{table}

\renewcommand{\baselinestretch}{1.2}
\begin{table}[htbp]
{\small
\caption{Results of the first simulation study with $\gamma=5/2$: comparison of existing interval designs with the proposed design based on $\text{GLR}_{\text{sd}}$ with $k_1=1.5$ and $k_2=1.05$ or 1.1, with respect to the accuracy of MTD estimation, the frequency of over-treatment, and the cost of the study (see section \ref{sim1} for details).}\label{sim.rst.3}
\newcolumntype{d}{D{.}{.}{1}}
\begin{center}
\begin{tabular}{cccccccccccccc}
\hline
\hline
\multicolumn{1}{c}{$D$}&\multicolumn{1}{c}{Design}&\multicolumn{1}{c}{$k_2$}&\multicolumn{3}{c}{$\phi=0.2$}&\multicolumn{1}{c}{}&\multicolumn{3}{c}{$\phi=0.25$}&\multicolumn{1}{c}{}&\multicolumn{3}{c}{$\phi=0.3$}\\
\cline{4-6}\cline{8-10}\cline{12-14}
\multicolumn{3}{c}{}&\multicolumn{1}{c}{\%MTD}&\multicolumn{1}{c}{\%OT}&\multicolumn{1}{c}{$N_{\text{ave}}$}&\multicolumn{1}{c}{}&\multicolumn{1}{c}{\%MTD}&\multicolumn{1}{c}{\%OT}&\multicolumn{1}{c}{$N_{\text{ave}}$}&\multicolumn{1}{c}{}&\multicolumn{1}{c}{\%MTD}&\multicolumn{1}{c}{\%OT}&\multicolumn{1}{c}{$N_{\text{ave}}$}\\
\hline
4&BOIN&&54.2&39.8&22.6&&57.8&40.3&23.2&&63.1&41.8&23.0\\
&TEQR&&53.4&39.3&22.6&&58.2&40.0&23.3&&63.4&42.2&23.0\\
&mTPI&&53.2&46.0&22.6&&59.6&43.1&23.3&&63.3&42.6&23.0\\
&i3+3&&54.5&43.1&22.7&&57.8&45.7&23.3&&63.2&42.4&23.0\\
&$\text{GLR}_{\text{sd}}$&1.05&53.8&36.3&22.5&&58.3&35.4&23.3&&62.7&36.1&22.9\\
&$\text{GLR}_{\text{sd}}$&1.1&54.1&36.9&22.7&&59.2&38.9&23.3&&62.3&36.9&23.0\\
\hline
6&BOIN&&46.8&32.8&34.7&&51.3&33.9&35.5&&55.8&36.7&35.3\\
&TEQR&&47.2&34.0&34.8&&50.3&34.0&35.4&&55.7&37.0&35.3\\
&mTPI&&46.7&40.3&34.8&&52.3&36.9&35.5&&55.8&36.8&35.2\\
&i3+3&&47.1&37.4&34.8&&51.1&40.1&35.5&&55.7&37.0&35.3\\
&$\text{GLR}_{\text{sd}}$&1.05&47.3&29.6&34.8&&51.8&28.0&35.5&&54.9&29.8&35.3\\
&$\text{GLR}_{\text{sd}}$&1.1&47.2&30.3&34.8&&51.8&31.8&35.5&&55.1&30.0&35.2\\
\hline
8&BOIN&&41.6&29.2&47.1&&45.2&31.1&47.7&&50.3&33.2&47.5\\
&TEQR&&41.1&30.6&47.0&&45.4&31.1&47.6&&50.8&33.5&47.4\\
&mTPI&&40.7&35.7&47.0&&46.6&33.3&47.6&&49.0&33.3&47.5\\
&i3+3&&41.3&33.9&47.1&&46.6&37.4&47.6&&49.7&33.7&47.5\\
&$\text{GLR}_{\text{sd}}$&1.05&41.5&25.0&47.0&&44.9&23.9&47.6&&49.7&25.2&47.4\\
&$\text{GLR}_{\text{sd}}$&1.1&41.4&25.9&47.0&&46.2&27.1&47.7&&49.7&25.7&47.5\\
\hline
\end{tabular}
\end{center}
}
\%MTD: percentage of trials that correctly find the MTD; \%OT: percentage of patients who are over-treated (i.e., treated at a dose above the true MTD); $N_{\text{ave}}$: average number of patients treated; BOIN: Bayesian optimal interval; TEQR: toxicity equivalence range; mTPI: modified toxicity probability interval.
\end{table}

\renewcommand{\baselinestretch}{1.05}
\begin{table}[htbp]
{\footnotesize
\caption{Results of the second simulation study: comparison of single-dose and joint likelihood designs based on $\text{GLR}_{\text{sd}}$, $\text{GLR}_{\text{iso}}$, $\text{GLR}_{\text{pow}}$ and $\text{GLR}_{\text{lgx}}$ with respect to the accuracy of MTD estimation, the frequency of over-treatment, and the cost of the study (see section \ref{sim2} for details).}\label{sim.rst.4}
\newcolumntype{d}{D{.}{.}{1}}
\begin{center}
\begin{tabular}{lclccccccccccc}
\hline
\hline
\multicolumn{1}{c}{Scenario}&\multicolumn{1}{c}{$D$}&\multicolumn{1}{c}{Design}&\multicolumn{3}{c}{$\phi=0.2$}&\multicolumn{1}{c}{}&\multicolumn{3}{c}{$\phi=0.25$}&\multicolumn{1}{c}{}&\multicolumn{3}{c}{$\phi=0.3$}\\
\cline{4-6}\cline{8-10}\cline{12-14}
\multicolumn{3}{c}{}&\multicolumn{1}{c}{\%MTD}&\multicolumn{1}{c}{\%OT}&\multicolumn{1}{c}{$N_{\text{ave}}$}&\multicolumn{1}{c}{}&\multicolumn{1}{c}{\%MTD}&\multicolumn{1}{c}{\%OT}&\multicolumn{1}{c}{$N_{\text{ave}}$}&\multicolumn{1}{c}{}&\multicolumn{1}{c}{\%MTD}&\multicolumn{1}{c}{\%OT}&\multicolumn{1}{c}{$N_{\text{ave}}$}\\
\hline
\multicolumn{14}{l}{Scenario 1 (consistent with the power model)}\\
&4&$\text{GLR}_{\text{sd}}$&74.0&27.5&19.1&&72.2&29.4&20.4&&71.1&29.8&20.3\\
&&$\text{GLR}_{\text{iso}}$&74.1&27.8&19.1&&72.0&29.6&20.4&&73.3&29.6&20.3\\
&&$\text{GLR}_{\text{pow}}$&77.0&29.4&19.0&&76.3&31.8&20.2&&73.8&31.0&20.4\\
&&$\text{GLR}_{\text{lgx}}$&72.6&27.9&19.2&&71.1&30.4&20.3&&71.0&29.9&20.3\\
\cline{2-14}
&6&$\text{GLR}_{\text{sd}}$&68.7&24.3&28.7&&66.9&24.8&30.3&&68.1&25.7&30.9\\
&&$\text{GLR}_{\text{iso}}$&68.7&23.9&28.8&&67.9&24.6&30.4&&67.3&25.2&30.3\\
&&$\text{GLR}_{\text{pow}}$&74.0&26.5&28.8&&72.0&29.0&30.5&&69.2&29.9&31.1\\
&&$\text{GLR}_{\text{lgx}}$&68.2&25.0&28.7&&66.8&26.5&30.3&&66.2&25.6&30.6\\
\cline{2-14}
&8&$\text{GLR}_{\text{sd}}$&65.8&21.4&38.1&&64.2&22.1&40.1&&63.5&22.3&40.4\\
&&$\text{GLR}_{\text{iso}}$&65.2&21.6&38.3&&63.8&22.1&40.6&&63.5&22.9&40.8\\
&&$\text{GLR}_{\text{pow}}$&71.4&25.0&38.4&&70.3&28.0&40.4&&68.5&27.5&40.6\\
&&$\text{GLR}_{\text{lgx}}$&64.9&22.4&38.1&&63.5&23.2&40.4&&62.4&21.9&40.9\\
\hline
\multicolumn{14}{l}{Scenario 2 (consistent with the logistic model)}\\
&4&$\text{GLR}_{\text{sd}}$&45.9&31.4&21.3&&45.2&30.8&22.5&&47.6&30.6&22.0\\
&&$\text{GLR}_{\text{iso}}$&44.9&31.1&21.2&&45.4&30.4&22.4&&48.4&30.0&21.8\\
&&$\text{GLR}_{\text{pow}}$&48.4&33.9&21.3&&49.0&33.8&22.5&&48.8&31.9&21.9\\
&&$\text{GLR}_{\text{lgx}}$&46.4&31.6&21.3&&47.3&31.4&22.4&&48.9&30.7&21.9\\
\cline{2-14}
&6&$\text{GLR}_{\text{sd}}$&38.9&27.5&31.1&&39.0&26.9&32.8&&41.6&26.4&32.0\\
&&$\text{GLR}_{\text{iso}}$&38.9&27.5&31.3&&39.6&27.0&32.8&&41.6&26.3&32.2\\
&&$\text{GLR}_{\text{pow}}$&42.8&30.1&31.1&&43.5&30.0&32.8&&43.2&28.2&32.0\\
&&$\text{GLR}_{\text{lgx}}$&41.3&27.7&31.2&&43.2&27.7&32.9&&44.1&26.6&32.0\\
\cline{2-14}
&8&$\text{GLR}_{\text{sd}}$&35.2&24.5&41.0&&36.3&25.8&42.5&&37.1&23.4&42.0\\
&&$\text{GLR}_{\text{iso}}$&34.0&24.7&41.1&&35.9&25.6&42.5&&37.2&23.9&42.2\\
&&$\text{GLR}_{\text{pow}}$&38.8&28.0&41.0&&41.6&29.8&42.9&&40.4&25.9&42.1\\
&&$\text{GLR}_{\text{lgx}}$&38.7&25.1&40.8&&42.4&26.9&42.9&&41.6&24.0&42.1\\
\hline
\multicolumn{14}{l}{Scenario 3 (consistent with neither model)}\\
&4&$\text{GLR}_{\text{sd}}$&46.8&30.8&23.1&&51.7&29.1&23.7&&54.7&31.1&23.4\\
&&$\text{GLR}_{\text{iso}}$&47.3&30.8&23.1&&52.1&29.2&23.6&&56.0&31.0&23.5\\
&&$\text{GLR}_{\text{pow}}$&39.7&36.3&23.1&&46.7&36.3&23.7&&50.1&35.0&23.3\\
&&$\text{GLR}_{\text{lgx}}$&45.0&31.6&23.0&&48.3&31.0&23.7&&50.9&30.8&23.5\\
\cline{2-14}
&6&$\text{GLR}_{\text{sd}}$&38.8&24.0&35.3&&42.3&22.4&35.7&&51.6&24.0&35.7\\
&&$\text{GLR}_{\text{iso}}$&39.6&24.8&35.2&&43.5&22.9&35.8&&48.2&25.1&35.6\\
&&$\text{GLR}_{\text{pow}}$&33.2&30.9&35.3&&38.4&32.0&35.8&&42.9&31.0&35.8\\
&&$\text{GLR}_{\text{lgx}}$&37.9&25.5&35.3&&41.5&24.6&35.8&&45.5&24.4&35.6\\
\cline{2-14}
&8&$\text{GLR}_{\text{sd}}$&33.0&20.7&47.4&&36.3&18.9&47.8&&41.9&19.8&47.9\\
&&$\text{GLR}_{\text{iso}}$&33.6&20.0&47.4&&36.5&18.5&47.8&&40.6&20.4&47.8\\
&&$\text{GLR}_{\text{pow}}$&28.3&28.3&47.4&&34.9&31.3&47.8&&37.7&32.7&47.9\\
&&$\text{GLR}_{\text{lgx}}$&31.2&21.5&47.4&&35.0&20.3&47.8&&38.0&20.5&47.7\\
\hline
\end{tabular}
\end{center}
}
\%MTD: percentage of trials that correctly find the MTD; \%OT: percentage of patients who are over-treated (i.e., treated at a dose above the true MTD); $N_{\text{ave}}$: average number of patients treated.
\end{table}

\end{document}